\documentclass[%
 reprint, %frontmatterverbose, %preprint, %preprintnumbers,
 nofootinbib, %nobibnotes, %bibnotes,
 amsmath,amssymb,
 aps, 
 showkeys,
 ]{revtex4-1}
\usepackage{hyperref}
\usepackage{color}
\usepackage[normalem]{ulem} %use sout for strikethrough

\newcommand{\mbf}{\boldsymbol}

\usepackage{diagbox}

\usepackage{supertabular}

\newcommand\TT{\rule{0pt}{2.6ex}}       
\newcommand\BB{\rule[-1.2ex]{0pt}{0pt}}

\newcommand{\bp}{\mbf{{p}}}
\newcommand{\bX}{\mbf{{q}}}
\newcommand{\bP}{\mbf{{k}}}
\newcommand{\brho}{\mbf{{\rho}}}
\newcommand{\ra}{\rangle}
\newcommand{\la}{\langle}
\newcommand{\hx}{{x}}
\newcommand{\hxi}{{x}_i}
\newcommand{\hxj}{{x}_j}
\newcommand{\hp}{{p}}
\newcommand{\hpi}{{p}_i}
\newcommand{\hpj}{{p}_j}
\newcommand{\hX}{{q}}
\newcommand{\hXi}{{q}_i}
\newcommand{\hXj}{{q}_j}
\newcommand{\hP}{{k}}
\newcommand{\hPi}{{k}_i}
\newcommand{\hPj}{{k}_j}
\newcommand{\hrho}{{\rho}}
\newcommand{\hrhoi}{{\rho}_i}
\newcommand{\hrhoj}{{\rho}_j}

\newcommand{\hL}{{L}}
\newcommand{\hlk}{\hL_k}
\newcommand{\ijk}{\varepsilon_{ijk}}

\newcommand{\PdotX}{\mbf{\hP}\cdot\mbf{\hX}}
\newcommand{\pdotx}{\mbf{\hp}\cdot\mbf{\hX}}
\newcommand{\sumi}{\displaystyle\sum_{i=1}^3}
\newcommand{\sumj}{\displaystyle\sum_{j=1}^3}
\newcommand{\sumk}{\displaystyle\sum_{k=1}^3}
\newcommand{\summ}[1]{\displaystyle\sum_{{#1}=0}^\infty}

\newcommand{\calp}{\mathcal{P}}
\newcommand{\calo}{\mathcal{O}}
\newcommand{\commij}{[\hxi,\hpj]}
\newcommand{\commxx}{[\hxi,\hxj]}

\newcommand{\ih}{i\hbar}
\newcommand{\dij}{\delta_{ij}}
\newcommand{\ie}{i.e., }
\newcommand{\eg}{e.g., }
\newcommand{\cf}{cf.\ }
\newcommand{\be}{\begin{equation}}
\newcommand{\ee}{\end{equation}}
\newcommand{\oo}[1]{\mathcal{O}({#1})}
\newcommand{\round}[1]{\left({#1}\right)}
\newcommand{\squ}[1]{\left[{#1}\right]}

\usepackage{soul}

\begin{document}

\preprint{APS/123-QED}

\title{A framework for nonrelativistic isotropic models based on generalized uncertainty principles}% Force line breaks with \\

\author{Andr\'e Herkenhoff Gomes}
\altaffiliation[Current address: ]{Departamento de F\'isica, Universidade Federal de Ouro Preto, Ouro Preto, MG, Brazil; \url{andre.gomes@ufop.edu.br.}}
%\email{}
\affiliation{%
 Departamento de Ci\^encias Naturais, Universidade Federal do Esp\'irito Santo, S\~ao Mateus, ES, Brazil%\\
}%


\begin{abstract}
The existence of a fundamental length scale in nature is a common prediction of distinct quantum gravity models. Discovery of such would profoundly change current knowledge of quantum phenomena and modifications to the Heisenberg uncertainty principle may be expected. Despite the attention given to this possibility in the past decades, there has been no common framework for a systematic investigation of so-called generalized uncertainty principles (GUP). In this work we provide such a framework in the context of nonrelativistic quantum mechanics. Our approach is based on very few assumptions: there is a fundamental length scale, space isotropy, invariance under parity and time reversal transformations, and symmetricity of the position and momentum operators. We show that simple dimensional analysis allows building a common framework for isotropic models based on GUP (we call those iGUP models). We discuss some popular GUP models in this context after elaborating on relevant theoretical aspects of the framework. At last, we translate current bounds on three often investigated GUP models into bounds on parameters of such common iGUP framework.
\end{abstract}


\maketitle

%%%%%%%%%%%%%%
\section{Introduction}
\label{sec:intro}
%%%%%%%%%%%%%%

The possible existence of a fundamental length in nature has been discussed since the early days of modern quantum theory itself \cite{hagar}. Most notably, in the 1930s and 1940s, Heisenberg advocated its existence as a cure for nonrenormalizable divergences in Fermi's theory for $\beta$-decay \cite{kragh}. As quantum field theory headed to full blossoming during the 1950s, and especially as Fermi's theory became better understood as an effective, low-energy limit of the now well-established electroweak theory, Heisenberg's motivation for a fundamental length gradually faded. On the other hand, there was also at that time speculation around quantum gravitational effects playing a role in the precision to which lengths could be measured. In particular, Mead's investigation on the effects of gravitation on the Heisenberg's microscope gedanken experiment came to the conclusion that not only gravitation modifies the Heisenberg's uncertainty principle (HUP) but also enforces a nonvanishing lower bound on the uncertainty on position measurements, then interpreted as associated with a fundamental length scale due to the universal coupling of gravitation to all known particles \cite{mead}. Up to the present, indications of a fundamental minimum length related to quantum gravitational phenomena sprang both while revisiting Heisenberg's gedanken experiment \cite{padmanabhan,adler-santiago} and from very different contexts such as string theories \cite{strings1,strings2}, lattice models \cite{greensite,scardigli2010}, gedanken experiments with black holes \cite{maggiore,scardigli}, loop quantum gravity \cite{rovelli,polymer}, noncommutative geometry \cite{NC,nc-gup}, and curved momentum-space \cite{curved-momentum} --- see \cite{hagar,garay,hossen2013} for a comprehensive review.

The introduction of a fundamental length appears then to be a desirable feature for a quantum theory of gravitation \cite{hagar} and the modification of the HUP a rather general and model-independent quantum gravitational prediction \cite{maggiore,scardigli}. One line of reasoning is the following \cite{maggiore}: The expression for the radius of a black hole should be taken as a definition in general relativity, not an experimentally testable prediction of the theory, as an observer has no direct access to it, but the situation changes when taking into account quantum gravitational phenomena. In particular, by detecting the Hawking radiation emitted by the black hole, an observed could trace it back to the black hole and determine its radius. Considering detection of photons of wavelength $\lambda$, there would be a Heisenberg's microscope-like quantum uncertainty $\Delta x^{(1)} \gtrsim \lambda$ and a quantum gravitational uncertainty $\Delta x^{(2)} \gtrsim 2G\Delta M/c^2 \sim \hbar G/c^3 \lambda = \ell^2_P/\lambda$, with $\ell_P$ the Planck length, due to the change on the black hole's mass, and therefore its radius, as part of the Hawking radiation process. Adding the two uncertainties linearly and rewriting in terms of the uncertainty $\Delta p$ on the black hole's $x$-momentum component, one arrives at
\be\label{black-hole-gup}
\Delta x \cdot \Delta p \gtrsim \hbar + \text{constant}\times\frac{\ell_P^2}{\hbar}(\Delta p)^2,
\ee
where the numerical constant is to be bounded by experimentation, and a nonvanishing lower bound $(\Delta x)_\text{min}$ on the uncertainty on position measurements is predicted to be proportional to $\ell_p\sim10^{-35}$ m. This relation is often interpreted as indeed a generalized uncertainty principle (GUP) that governs all measurement processes in quantum gravity for three main reasons \cite{scardigli}, namely, (1) it implies a nonvanishing minimum observable length; (2) there is no dependence on the charge, mass or angular momentum of the black hole, therefore no memory of the particular system it refers to; and (3) it is also a result of gedanken experiments in the context of string theories. Within such interpretation, relations as this one have been used to investigate the effects of a fundamental length scale in a broad variety of situations, ranging from astrophysical and cosmological context to microscopic quantum mechanical systems \cite{hossen2013,tawfik-diab-2014,tawfik-diab-2015}, and nonlinear extensions of the Schr\"odinger equation within the exact uncertainty principle approach \cite{nonlinear1,nonlinear2}.

There is an arbitrariness on the step of linearly adding uncertainties to get the above GUP nevertheless, and without a definite theory for quantum gravity there is no clear physical principle guiding what features a GUP should actually have besides reducing to the HUP in the low energy limit. A common approach, and the one we consider in this work, takes any deviation from the HUP as a universal kinematical effect coming from modifications on the algebra of the position and momentum operators \cite{maggiore-algebra}. In particular, the uncertainty relation between the two operators is obtained from the well-known inequality $\Delta x_i\cdot\Delta p_j \ge \frac{1}{2}|\la\commij\ra|$, where $\hxi$ and $\hpi$ stand for the components of the position and momentum operators, respectively, where $i=\{1,2,3\}$ \cite{cohen1}. Deviations from the HUP can be studied starting from a generalization of the canonical commutation relation,
\be\label{starting-point}
\commij = \ih f_{ij}(\hx,\hp;\dots\text{?}),
\ee
with $f_{ij}$ some function of the position and momentum operators and parameters related to yet unknown degrees of freedom from physics beyond the standard model. Lacking a clear guiding principle for a definite form for $f_{ij}$, considerable attention has been given to proposals embracing predictions of a nonvanishing minimum uncertainty on position measurements \cite{kmm95, kempf97, kempf-mangano97, nouicer07,ali-das-vagenas09, pedram12-plb2, chung-hass2019}. The arguably most investigated \cite{kmm95} assumes $f_{ij} = \dij(1+\beta\hp^2)$, leading to $(\Delta x_i)_\text{min}=\hbar\sqrt{\beta d}$ for $d$-dimensional space. Most experimental searches for evidences of a minimal position uncertainty revolve around this proposal and set bounds on $\beta$, often rewritten as $\beta_0 \ell^2_P/\hbar^2$, where $\beta_0$ is dimensionless, expected to be of order 1 \cite{scardagli-lambiase-vagenas}, and currently reported as smaller than $5.2 \times 10^{6}$ \cite{bushev} --- though this bound depends on subtle assumptions and should be taken with care (see Section \ref{sec:bounds-igup}).

Previous works proposing different GUPs usually share the same following programme. A particular $f_{ij}$ is chosen, but the motivation for its expression varies widely. Then theoretical investigation of the proposed commutator proceeds along with derivation of physical predictions. This leads to increasing understanding, in a broad range of systems, of the consequences of modifying the HUP and also guides experimentalists to test new predictions. Nevertheless, there was never a systematic approach to different GUP proposals. There is no common framework connecting the different proposals nor the different parameters appearing within each proposal. As a consequence,
\begin{itemize}
\item as pointed out by Hossenfelder back in 2013 \cite{hossen2013}, this makes it harder to establish a compilation of bounds on GUPs;
\item experimental bounds are often placed on parameters associated to some specific one-dimensional GUP model, but it is not clear the physical significance of such model as it can be derived from very different three-dimensional models; and
\item to further complicate the situation, as many of the different GUP proposals share the same symbol $\beta$ for their basic parameter, this oversimplification threatens to confuse bounds on parameters from one proposal for bounds on a totally different proposal --- something already present in the literature.
\end{itemize}

The goal of this work is then twofold: (1) to provide a common framework for the class of nonrelativistic GUP models derived from rotationally covariant $f_{ij}$ and (2) to summarize in this context often cited bounds on GUP parameters. This framework is expected to encompass all nonrelativistic isotropic models based on GUP (iGUP models), each one corresponding to a specific set of parameters present in the framework. The novelty of our approach is that instead of starting from a modification of the canonical commutator relation, we \textit{derive} its general nonrelativistic isotropic extension after \textit{constructing} operators $\hxi$ and $\hpi$ based on the simple assumption that a fundamental length scale exists in nature and a few other reasonable technical requirements.

The rest of this paper is organized as follows. Section \ref{sec:construction-of-iGUP} is devoted to the construction and discussion of the iGUP framework. In Sec.\ \ref{sec:applications} we discuss some representative GUP models proposed in the literature in the context of the framework we devised. Current experimental bounds on GUP models are translated into bounds on iGUP parameters in Sec.\ \ref{sec:bounds-igup}. We close this paper with our final remarks in Sec.\ \ref{sec:conclusions}.

%%%%%%%%%%%%%%
\section{Construction of the iGUP framework}
\label{sec:construction-of-iGUP}
%%%%%%%%%%%%%%

A common framework for GUP models is currently not available. The reason may be a considerable number of works on the field take the commutator (\ref{starting-point}) as the starting point of investigation, each with $f_{ij}$ chosen according to diverse motivations.

One apparent common ground is the interpretation of $\hxi$ and $\hpi$ as \textit{physical} operators associated with position and momentum components satisfying unconventional canonical commutation relation $\commij=\ih f_{ij}$. Momenta are kept as usual, $[p_i,p_j]=0$, but, in accordance to the Jacobi identity, position operators may be noncommutative, $\commxx\neq0$. To overcome the superior complexity of such algebra, the strategy is then using a representation of $x_i$ and $p_i$ in terms of a pair of \textit{auxiliary} operators, here denoted by $q_i$ and $k_j$, satisfying the conventional algebra $[\hXi,\hPj]=\ih\dij$ and $[\hXi,\hXj]=[\hPi,\hPj]=0$.\footnote{Auxiliary operators are sometimes interpreted as low energy position and momentum operators, but we believe this interpretation is error-prone specially for commutative GUP models (\cf Sec.\ \ref{sec:P=p}); thus we avoid it completely.} Physical predictions --- \eg in the form of low energy perturbations to the ground state energy of a system --- can then be obtained once a representation $\hxi=\hxi(\bX,\bP)$ and $\hpi=\hpi(\bX,\bP)$ satisfying $\commij=\ih f_{ij}$ is chosen. Despite the operational advantages of this approach, it does not provide a common language embracing and connecting different proposals for $f_{ij}$.

In this section we fill this gap constructing a common framework for isotropic models based on GUP, the iGUP framework. Our approach differs from the described above in the sense we first \textit{derive} the operators $\hxi(\bX,\bP)$ and $\hpi(\bX,\bP)$ based only on few basic requirements; these are
\begin{itemize}
    \item spatial isotropy,
    \item existence of a fundamental length scale,
    \item conventional behavior under parity and time reversal transformations, and
    \item the operators $x_i$ and $p_i$ are symmetric.
\end{itemize}
Next, we use spatial isotropy to write the allowed form of position and momentum operators, $x_i$ and $p_i$, as functions of the auxiliary operators, $q_i$ and $k_i$. Then, from the assumption that a fundamental length scale exists, dimensional arguments impose severe restrictions on the form of $x_i$ and $p_i$. Demanding both operators have conventional behavior under parity and time reversal transformation further restricts functions of $q_i$ and $k_i$ to be real or imaginary. At last, demanding symmetricity of $x_i$ and $p_i$ enforces a final general expression for these operators.

Notice the only unconventional assumption we set forth is the existence of a fundamental length scale. As we see next, this is enough to provide expressions for $x_i$ and $p_i$ leading to an extension $\commij = \ih\dij \to \ih f_{ij}$ that is general enough to encompass all GUP models formulated in isotropic space.

%%%%%%%%%%%%%%
\subsection{Position and momentum operators}
\label{sec:isotropic-operators}
%%%%%%%%%%%%%%

The first step in our approach is writing $\hxi$ and $\hpi$ as general \textit{isotropic} combinations of $\hXi$ and $\hPi$ \cite{wang}:
\be\label{expansion-3d-x}
\hxi = F(\hX,\hP,\PdotX)\hXi + \hPi G(\hX,\hP,\PdotX),
\ee
\be\label{expansion-3d-p}
\hpi = H(\hX,\hP,\PdotX)\hPi + \hXi I(\hX,\hP,\PdotX),
\ee
where $F$, $G$, $H$, and $I$ are functions of $\hX\equiv|\mbf{\hX}|$, $\hP\equiv|\mbf{\hP}|$ and $\PdotX$.\footnote{The ordering of auxiliary operators $\hXi$ and $\hPi$ --- \eg as in $\PdotX$ --- is irrelevant for the arguments in the main text. In the end, if a different ordering is chosen, we arrive at an expression for $\hxi$ that is different only up to canonical transformations. In this sense, the ordering of the auxiliary operators is immaterial.} Our \textit{basic hypothesis} is that these functions also depend on parameters possibly related to a fundamental length scale and which we generically denote by $\alpha$, where $[\alpha]=[p]^{-1}$ --- \eg function $F$ may depend, among others, on a parameter of dimension $[\alpha]$ while function $G$ may depend on a different parameter of dimension $[\alpha]^2$. At first, the proposed dimensionality of $\alpha$ may sound odd as one could expect it to have dimension of length instead, but taking any nonvanishing minimum position uncertainty as depending on positive powers of both $\hbar$ and $\alpha$, simple dimensional analysis reveals $[\alpha]=[p]^{-1}$ indeed. 

The next step it to use \textit{dimensional analysis} to engineer the functions $F$, $G$, $H$, and $I$ on the basis of four main requirements:
\begin{itemize}
    \item[(i)] none of them have dimensional dependence on inverse powers of $[\hbar]$ or $[\alpha]$ to ensure new physics is relevant only at higher energies;
    \item[(ii)] conventional physics is recovered for $\alpha\to0$, so these functions must have dimensional dependence on $[\alpha]$ at least; for the same reason, 
    \item[(iii)] $\hxi$ and $\hpi$ are finite at $\hX\to0$ and $\hP\to0$; and, in particular,
    \item[(iv)] $\hxi$ and $\hpi$ reduces to $\hXi$ and $\hPi$, respectively, at the $\hP\to0$ limit.
\end{itemize}
To begin with, function $F$ is dimensionless, so its structure is constrained to the dimensionless combinations of $[q]$, $[k]$, $[\hbar]$ and $[\alpha]$,
\be
\frac{[q][k]}{[\hbar]}, \quad
\frac{[\hbar]}{[q][k]},
\quad
[\alpha][k],
\quad
\frac{1}{[\alpha][k]}, \quad
\frac{[q]}{[\alpha][\hbar]},
\quad
\frac{[\hbar][\alpha]}{[q]},
\ee
but the only satisfying requirements (i) and (ii) are $[\alpha][k]$ and $[\hbar][\alpha]/[q]$. Both satisfy (iii) because $F$ is accompanied by $\hXi$ in (\ref{expansion-3d-x}), but only $[\alpha][k]$ and any positive power of it satisfy (iv). This amounts to writing $F$ as a simpler function $f(k)$ given by
\be\label{functions-f}
f(\hP) = \summ{n} a_n \hP^n
\quad \text{with} \quad
[a_n] = [\alpha]^n,
\ee
with $a_0\equiv1$ to recover standard quantum mechanics when parameters of dimension $[\alpha]^n$ are set to zero.\footnote{Another valid choice is $a_0$ depending on $\alpha$ such that $a_0\to1$ for $\alpha\to0$; for instance, $a_0=1+(\text{const.})\times\alpha$. Dimensional analysis of this simple case reveals the constant depends on the particle's mass, suggesting phenomena originated by the existence of a fundamental length would be mass-dependent \cite{maggiore-algebra,maggiore2021}. We will not consider this possibility here, but it can be completely accommodated in our approach by setting $a_0\neq1$ instead.} The same argument is valid for the also dimensionless function $H$, which we rewrite as $h(k)$ given by
\be\label{functions-h}
h(\hP) = \summ{n} \alpha_n \hP^n
\quad \text{with} \quad
[\alpha_n] = [\alpha]^n,
\ee
with $\alpha_0\equiv1$ for the same reason as that for $a_0$. Moving on to function $G$, its dimension $[q]/[k]$ can be expressed as
\be\label{dimensions-G}
[\alpha][q],
\quad
\frac{[q]^2}{[\hbar]},
\quad
\frac{[\hbar]}{[k]^2},
\quad
\frac{[\alpha][\hbar]}{[k]},
\quad
[\alpha]^2[\hbar],
\ee
but the only acceptable on the grounds of the four criteria are $[\alpha][q]$, $[\hbar][\alpha]/[k]$, and $[\hbar][\alpha]^2$, which may also be accompanied by any power of the dimensionless $[\alpha][k]$. As a result, function $G$ is the sum of three functions: one of dimension expressed as $[\alpha][q]$,
\be\label{functions-c}
c(\hP)\hX = \sum_{n=0}^\infty c_n\hP^n\hX
\quad \text{with} \quad
[c_n] = [\alpha]^{n+1};
\ee
another complying with both $[\hbar][\alpha]/[k]$ and $[\hbar][\alpha]^2$,
\be\label{functions-gamma}
\ih\gamma(\hP) = \ih\sum_{n=-1}^\infty \gamma_n\hP^n
\quad \text{with} \quad
[\gamma_n] = [\alpha]^{n+2},
\ee
where the factor of $i$ is introduced for later convenience and the contribution $\hP^{-1}$ respects requirement (iii) because $G$ is accompanied by $\hPi$ in (\ref{expansion-3d-x}); and another function also based on $[\hbar][\alpha]/[k]$ and $[\hbar][\alpha]^2$ but after considering $[\hbar]=[q][k]$,
\be\label{functions-g}
g(\hP)(\PdotX) = \sum_{n=-1}^\infty b_n\hP^n (\PdotX)
\quad \text{with} \quad
[b_n] = [\alpha]^{n+2}.
\ee
At last, function $I$ in (\ref{expansion-3d-p}) has dimension $[k]/[q]$, which can be expressed as the inverse of any of those listed in (\ref{dimensions-G}), but neither satisfy criteria (i); hence, $I=0$.

From the above, operators $\hxi$ and $\hpi$ take the temporary form $\hxi = f(\hP)\hXi +g(\hP)\hPi(\PdotX) +\ih\gamma(\hP)\hPi +c(\hP)\hPi\hX$, and $\hpi=h(\hP)\hPi$. This is as far as we can go with arguments based only on spatial isotropy and dimensional constraints, but we already note these arguments are sufficient to enforce commutativity of the momentum operator components,
\be
[\hpi,\hpj]=0.
\ee
A further step is to demand conventional behavior under \textit{parity} transformation, where
\be
\mbf{\hx}\to-\mbf{\hx}
\quad \text{and} \quad \mbf{\hp}\to-\mbf{\hp},
\ee
as well as under \textit{time reversal} transformation, where
\be
\mbf{\hx}\to\mbf{\hx},
\quad
\mbf{\hp}\to-\mbf{\hp},
\quad \text{and} \quad
i \to -i.
\ee
These requirements enforce that the coefficients $a_n$, $b_n$, $\gamma_n$, and $\alpha_n$ are real but $c_n$ is purely imaginary.

The final ingredient for our construction is requiring $\hxi$ to be a \textit{symmetric} operator, that is, $\la\psi|\hxi|\phi\ra = \la\phi|\hxi|\psi\ra^\ast$. This condition ensures $\hxi$ has real spectrum even though a representation on its basis may not be achievable (\eg due to a fundamental length scale) \cite{kmm95,kempf2000,pedram12-prd,pasquale-xbasis}. In the general case, $\hxi$ is symmetric under some definition for the scalar product as long as $\gamma(\hP)$ is \textit{non observable} and can be suitably adjusted, a freedom that exists only if both $\commij$ and $\commxx$ have no dependence on it, but this is true only if $c(\hP)$ vanishes. Hence,
\be
\hxi \quad \text{symmetric} \quad \iff \quad c(\hP)=0.
\ee
This conclusion is in harmony with explicit calculations whose results we present in the next sections. In passing, we notice symmetricity (and actually hermiticity) of $\hpi$ is a natural feature of the model.

To close this section, we summarize our results with the final form of $\hxi$ and $\hpi$,
\be\label{operator-x-general}
\hxi = f(\hP)\hXi +g(\hP)\hPi(\PdotX) +\ih\gamma(\hP)\hPi
\ee
\be\label{operator-p-general}
\quad \hpi=h(\hP) \hPi.
\ee
Each function $f$, $g$, $\gamma$, and $h$ contains a set of coefficients corresponding to the respective power series on $\hP$ discussed before. In particular,
coefficients $a_n$ and $\alpha_n$ have dimension $[\alpha]=[p]^{-1}$, and $b_n$ has dimension $[\alpha]^{n+2}$; these are real and observable, and from here and on will be referred to as the iGUP parameters. Coefficients $\gamma_n$ are real but unobservable, to be adjusted to ensure $\hxi$ is a symmetric operator (\cf Sec.\ \ref{sec:P=p}). As a final remark, notice although $g$ and $\gamma$ may develop poles at $\hP\to0$, the operator $\hxi$ remains perfectly regular and reduces to $\hXi$ in this limit.

%%%%%%%%%%%%%%
\subsection{Commutators}
\label{sec:commutators}
%%%%%%%%%%%%%%

Straightforward calculation reveals the commutator of $\hxi$ and $\hpi$ assumes the very general form
\be\label{commutator-xp}
\commij = \ih\Big\{ fh\dij +\Big[ \frac{fh'}{\hP} +g(h+\hP h')
\Big]\hPi\hPj \Big\},
\ee
where primes represent derivatives with respect to $\hP$ --- \eg $f'\equiv df/d\hP$. Notice the factor $1/\hP$ may pose no issue due to the factor $\hPi\hPj$. A word of caution regarding this result is that the right-hand side above depends on auxiliary operators and does \textit{not} really offer any detailed glimpse on how $\commij$ depends upon the physical position and momentum operators. We deal with this limitation in Secs.\ \ref{sec:P=p} and \ref{sec:X=x}. The same remark is valid for the commutator of position operators,
\be\label{commutator-xx}
\commxx = \ih\Big[ g \Big( f-\hP f' \Big) - \frac{ff'}{\hP} \Big]\sumk\ijk\hlk,
\ee
where the operator 
\be\label{operator-L-aux}
\hL_k\equiv\sumi\sumj \varepsilon_{ijk}\hXi\hPj
\ee
is not, at this moment, straightforwardly interpreted as the generator of rotations --- see next section. Similar to (\ref{commutator-xp}), we expect the inverse dependence on $\hP$ to be innocuous due to the compensating factor of $\hPi$ provided by $\hlk$. Although the right-hand side on (\ref{commutator-xx}) also depends on auxiliary operators, it already signals a possible nonvanishing uncertainty on successive measurements of different spatial coordinates because the components $\hxi$ are generally noncommutative. It also reveals specific commutative models can be devised once functions $f$ and $g$ satisfy a \textit{commutativity condition},
\be\label{commutativity-condition}
g=\frac{ff'}{\hP (f-\hP f')}.
\ee
Since here $f$ and $g$ are functions of $\hP$, this condition is somewhat more general than the one found in the literature \cite{kempf97,kempf-mangano97}; namely, $g=2ff'/(f-2\hP^2f')$ with primes there denoting $d^2/d\hP^2$, and $f=f(\hP^2)$ and $g=g(\hP^2)$.

The framework based on the set of equations (\ref{operator-x-general})--(\ref{commutator-xx}) leads to what will be called here \textit{isotropic generalized uncertainty principle} models or just iGUP models for short. Those additionally satisfying the commutativity condition (\ref{commutativity-condition}) will be referred to as commutative iGUP models. Notice iGUP models based on different choices for $\hxi$ and $\hpi$ are equivalent as long as these are related by isometric canonical transformations, which preserve the commutator algebra \cite{anderson1,anderson2} --- \cf \cite{kmm95,kempf97,kempf-mangano97,pedram12-prd} for specific cases. 
Investigation of any particular iGUP model may be facilitated by the \textit{choice} of operator $\hxi$ as in (\ref{operator-x-general}) but with $\hpi=\hPi$, further simplifying the momentum space representation of the wavefunction because $[\hXi,\hpj]=\ih\dij$ in this case. This approach is discussed in the next section. 
For \textit{commutative} models, and only for these, an alternative choice is also available setting $\hxi=\hXi$ while keeping $\hpi$ as in (\ref{operator-p-general}) and interpreting $\hPi$ as the generator of translations $\hrhoi$, such that $[\hxi,\hrhoj]=\ih\dij$, even though position space representation of the wavefunction may not be achievable. We discuss this alternative choice for commutative models in Sec.\ \ref{sec:X=x} and its equivalence to the other approach in Sec.\ \ref{sec:equivalent-models}.

%%%%%%%%%%%%%%
\subsection{Generally noncommutative iGUP models with \texorpdfstring{$\mbf{\hpi=\hPi}$}{}}
\label{sec:P=p}
%%%%%%%%%%%%%%

Since any iGUP model has commutative momentum components, it is in principle always possible to find unitary transformations on $\hxi$ and $\hpi$ to rescale the auxiliary operator $\hPi$ by a factor of $1/h$ and make it match $\hpi$. In this case, $\hxi$ is given by
\be\label{P=p:operator-x}
\hxi = f(\hp)\hXi +g(\hp)\hpi(\pdotx) +\ih\gamma(\hp)\hpi,
\ee
and $[\hXi,\hpi]=\ih\dij$. The general expression (\ref{commutator-xp}) for the commutator of the position and momentum operators simplifies to
\be\label{P=p:commutator-xp}
\commij=\ih ( f\dij + g\hpi\hpj ),
\ee
and is now expressed entirely in terms of physical operators. From the relation $\Delta\hxi\cdot\Delta\hpi\ge\frac{1}{2}|\la\commij\ra|$, we see that specific choices for $f$ and $g$ may lead to nonvanishing minimum position uncertainty although, as far as we know, there is no general conclusion on what exact properties $f$ and $g$ must possess to do so. Some models with this feature are discussed in Sec.\ \ref{sec:applications} in connection to our framework.

The commutator for components of the position operator takes the form
\be\label{P=p:commutator-xx}
\commxx = \ih\Big[ g\Big(f-\hp f'\Big) - \frac{ff'}{\hp} \Big]\sumk\ijk\hlk,
\ee
and is also entirely expressed as a function of $\hxi$ and $\hpi$ since
\be\label{P=p:angular-momentum}
\hL_k \equiv \sumi\sumj \varepsilon_{ijk}\hXi\hPj = \frac{1}{f(\hp)} \sumi\sumj \varepsilon_{ijk}\hxi\hpj.
\ee
As long as spin is neglected, the operator $L_k$ can be identified as the orbital angular momentum \cite{maggiore2021}, acting as the generator of rotations implemented by
\be
U(\mbf{\theta}) = \exp\{ -i \mbf{\theta}\cdot\mbf{\hL}/\hbar\},
\ee
where $\mbf{\theta}=\theta\mbf{\hat{n}}$ with $\mbf{\hat{n}}$ is the unit vector along the rotation axis and $\theta$ the angle of rotation around this axis. Notice $\hL_k$ satisfies the conventional rotation algebra,
\be
[\hpi,\hL_j] = \ih \sumk \varepsilon_{ijk}\hp_k,
\ee
\be
[\hxi,\hL_j] = \ih \sumk \varepsilon_{ijk}\hx_k,
\ee
\be
[\hL_i,\hL_j]=\ih \sumk \varepsilon_{ijk}\hL_k,
\ee
implying, for instance, the operators $\hL^2$ and $\hL_3$
have eigenvalues $\ell(\ell+1)\hbar^2$ and $m_\ell\hbar$, respectively, associated to a common set of eigenvectors $|\ell,m_\ell\ra$ with quantum numbers $\ell=0,1,2,\dots$ and $m_\ell = -\ell,-\ell+1,\dots,\ell-1,\ell$, as in standard quantum mechanics.

At this point we should mention expressions (\ref{P=p:commutator-xp}) and (\ref{P=p:commutator-xx}) for the commutators, as well as their discussed consequences, are not new in the literature --- \eg see \cite{kempf97,kempf-mangano97} --- but they indicate the consistency of our approach so far. On the other hand, taking advantage of our construction in Sec.\ \ref{sec:isotropic-operators}, these two commutators can be expressed as power series on the physical momentum,
\be\label{P=p:commutator-xp-power-series}
\commij = \ih \Big[ \dij + \sum_{\calp=1}^\infty \Big( a_{\calp}\dij +b_{\calp-2} \frac{\hpi\hpj}{\hp^2} \Big) \hp^{\calp} \Big],
\ee
and
\begin{align}\label{P=p:commutator-xx-power-series}
\commxx = -\ih \sum_{\calp=0}^\infty \sum_{n=0}^{\calp} & a_n \Big[ (\calp-n+1) a_{\calp-n+1} \nonumber\\
& + (n-1)b_{\calp-n-1} \Big] \hp^{\calp-1} \ijk\hlk,
\end{align}
which may be useful for approximate calculations. For instance, in the particular case of \textit{commutative} models, coefficients of $f(\hp)$ and $g(\hp)$ necessarily satisfy $a_1=b_{-1}$ and $2a_2+a_1^2=b_0$, meaning that unless $g$ is singular at $p=0$ the first iGUP correction to $\commij$ comes only at $\oo{\alpha^2}$. Having the above power series for the commutators is a distinguishing feature of our approach as it derives from the expressions constructed for $f$ and $g$ in Sec.\ \ref{sec:isotropic-operators} based on rather general arguments.

Since physical momentum operators $\hpi$ are commutative and have no associated nonvanishing minimum uncertainty, a Hilbert space representation on momentum wavefunctions can be found with $\tilde{\psi}(\mbf{p}) \equiv \la\mbf{p}|\psi\ra$ for $\hpi|\mbf{p}\ra=p_i|\mbf{p}\ra$. Then, commutators (\ref{P=p:commutator-xp}) and (\ref{P=p:commutator-xx}) are realized for the $\hpi$ acting as the conventional multiplicative operator and $\hxi$ as a modified derivative operator,
%
%\be
%\hpi \psi(\mbf{p}) = p_i \psi(\mbf{p}),
%\ee
%
\be
\hxi \psi(\mbf{p}) = \ih \left[ f(p) \partial_{p_i} + g(p) p_i (\mbf{p} \cdot \partial_{\mbf{p}} ) + \gamma(p)p_i \right]  \psi (\mbf{p}),
\ee
The function $\gamma(p)$ is arbitrary, unobservable as it does not affect the expressions for $\commxx$ and $\commij$; thus, we choose it so as to enforce that $\hxi$ is a symmetric operator under some definition of the scalar product. In particular, $\la\psi|\hxi|\phi\ra = \la\phi|\hxi|\psi\ra^\ast$ is satisfied for the scalar product defined as
\begin{align}\label{P=p:scalar-product}
\la\psi|\phi\ra = \int \frac{d^3p}{\round{f + g p^2}^{1-\varepsilon}} &  \psi^\star(\mbf{p})\phi(\mbf{p}), \nonumber\\
& \!\!\!\!\!\!\!\! \text{where} \quad
\varepsilon = \frac{2p(\gamma-g)}{f'+g'p^2+2pg},
\end{align}
which fixes $\gamma(p)$ for any desired $\varepsilon$. From this definition it also follows the operator $\hpi$ is not only symmetric, but also hermitian since its domain matches with that of its adjoint. Although rendered a symmetric operator, then possessing only real eigenvalues \cite{kempf2000}, $\hxi$ may not be hermitian depending on whether functions $f$ and $g$ lead or not to a model with minimal uncertainty on position \cite{kmm95}. If they do, $\hxi$ is symmetric but not hermitian and, therefore, a position representation for the wavefunction is \textit{not} achievable as there is no orthonormal set of eigenvector for the set of real eigenvalues of $\hxi$. Overlooking technical details, the core idea revolves around that, by definition, position eigenvectors $|x_i\ra$ have vanishing position uncertainty, $\la x_i|(\Delta\hxi)^2|x_i\ra\equiv0$, and thus cannot be approximated by any physical state $|\psi_\text{phys}\ra$ for models where 
$\la\psi_\text{phys}|(\Delta\hxi)^2|\psi_\text{phys}\ra \ge (\Delta xi)^2_\text{min}>0$. Further detailed functional analysis for any particular iGUP model with $\hxi(\bX,\bp)$ and $[\hXi,\hpj]=\ih\dij$ can be done, for instance, following the recipe of Refs.\ \cite{kmm95,pedram12-prd,kempf-mangano97}. Information regarding the spatial localization can be recovered only to a limited extent; \eg using maximally localized quantum states \cite{kmm95,detournay,bernardo-esguerra,pasquale-xbasis}.

%%%%%%%%%%%%%%
\subsection{Commutative iGUP models with \texorpdfstring{$\mbf{\hxi=\ih\partial_{p_i}}$}{}}
\label{sec:X=x}
%%%%%%%%%%%%%%

A remarkable consequence of $\commxx\neq0$ is that translation invariance is no longer a fundamental symmetry of the space. This may be an undesirable feature as conservation of momentum is no longer ensured. In this context, here we restrict ourselves to the particular class of commutative iGUP models, \ie those satisfying
\be\label{X=x:comm-cond}
\commxx=0
\quad \Longleftrightarrow \quad
g=\frac{ff'}{\hp (f-\hp f')},
\ee
where, as in the previous section, Hilbert space representation on momentum wavefunctions is used to express the commutativity condition (\ref{commutativity-condition}).

Restriction to commutative models also allows for a Hilbert space representation on wavefunctions of the generator of \textit{translations} $\hrhoi$. The reason is that, for the operator 
\be
U(\mbf{d}) = \exp\{ -i \mbf{d}\cdot\brho/\hbar \}
\ee
implementing translations of $\mbf{d}$ as $U(\mbf{d})|\mbf{x}\ra=|\mbf{x}+\mbf{d}\ra$ and $U(\mbf{d})\mbf{\hx}U(\mbf{d})^\dagger = \mbf{\hx}-\mbf{d}$, there is $\hrhoi$ satisfying 
\be
[\hxi,\hrhoj]=\ih\dij
\ee
and acting as the generator of such translations. Straightforward substitution of (\ref{P=p:operator-x}), along with (\ref{X=x:comm-cond}) to relate $f$ and $g$, reveals the generator of translations $\hrhoi$ and the momentum operator $\hpi$ are \textit{not} the same, but are related by
\be\label{X=x:operator-P}
\hrhoi = \frac{\hpi}{f(\hp)}
\quad \text{or} \quad 
\hpi = h(\hrho) \hrhoi.
\ee
The identification $h(\hrho) \equiv f(\hp(\hrho))$ suggests the auxiliary operator $\hPi$ in (\ref{operator-p-general}) corresponds to the generator of translations $\hrhoi$, but only within this representation; thus, we retain the symbol $\hrhoi$ only in this case to make distinction of its meaning clear. Wavefunctions on $\rho$-space are $\psi(\mbf{\rho})\equiv\la\psi|\mbf{\rho}\ra$ where $\hrhoi|\mbf{\rho}\ra=\rho_i|\mbf{\rho}\ra$, and commutators (\ref{P=p:commutator-xp}) and (\ref{P=p:commutator-xx}) are then realized for
\be
\hpi \psi(\mbf{\rho}) = h(\rho)\hrhoi \psi (\mbf{\rho}),
\ee
\be
\hxi \psi(\mbf{\rho}) = \ih \partial_{\rho_i} \psi (\mbf{\rho}).
\ee
Even though the position operator looks conventional, $\hxi = \ih \partial_{p_i}$, it may not be hermitian despite being symmetric under the scalar product
\be\label{X=x:scalar-product}
\la\psi|\phi\ra = \int_{I} d^3\rho\, \psi^\star(\mbf{\rho})\phi(\mbf{\rho})
\ee
within the integration range $\rho_i \in I$ for which $h(\rho)$ is defined. The technical reason is that $\hxi$ may not be a symmetric operator on the exact same domain as its adjoint if $I$ does not cover the whole range of real numbers \cite{pedram12-prd,kempf-mangano97}. The physical reason, as discussed by the end of the previous section, is that hermiticity of $\hxi$ depends on whether there is a nonvanishing minimum position uncertainty --- and at least for one-dimensional models, this is known to depend on $I$ \cite{mangano2016}. Both reasons now relate to $h(\hrho)$ since it determines both the integration range $I$ in (\ref{X=x:scalar-product}) and the structure of $\commij$ in (\ref{commutator-xp}).

Investigation of any commutative iGUP model within the choice $\hxi$ and $\hpi(\brho)$ with $[\hxi,\hrhoj]=\ih\dij$ as above demands an expression for $\commij$ as function of the physical momentum operator $\hpi$ only. In what follows we derive this expression and the result is given by (\ref{X=x:commutator-xp}). Since commutative models are very popular in the literature, this result plays an important role to the rest of this paper. We use it to discuss equivalent operator representations for commutative models and the relation of three- and one-dimensional models (Secs.\ \ref{sec:equivalent-models} and \ref{sec:3d-extension-from-1d}); to place some commutative GUP models of the literature in the iGUP framework (Sec.\ \ref{sec:applications}); and in the translation of bounds on commutative models into bounds on iGUP parameters (Sec.\ \ref{sec:bounds-igup}).

Position and momentum commutation relation (\ref{commutator-xp}) for $\hxi=\hXi$ and $\hpi=h(\hrho)\hrhoi$ simplifies to
\begin{align}\label{X=x:commutator-xp-low-p}
\commij & = \ih\round{ h\dij +\frac{h'}{\hrho}\hrhoi\hrhoj } \nonumber\\
& = \ih\summ{n} \alpha_n \left( \dij + n \frac{\hrhoi\hrhoj}{\hrho^2} \right) \hrho^n.
\end{align}
Expressing the right-hand side in terms of the momentum $\hp$ instead of the generator of translations $\hrho$ requires, as a first step, the use of series reversion on (\ref{X=x:operator-P}) to write $\hrhoi$ as a function of $\hpi$, 
\be\label{X=x:lagrange-inversion}
\hrhoi = \hpi \sum_{r=0}^\infty A_r \hp^r,
\ee
where the coefficients $A_r$ are found from the Lagrange inversion theorem, here leading to
\be
A_r = \frac{1}{(r+1)!} \lim_{\rho\to0} \frac{d^{r}}{d\rho^{r}} [h(\rho)]^{-r-1},
\ee
or more efficiently from \cite{chang1987},
\be
A_r = \sum_{s=0}^r \begin{pmatrix} r+s+1 \\ s \end{pmatrix} \frac{(-1)^s}{r+s+1} q_{r,s},
\ee
where
\be
q_{r,s} = \sum_{i=1}^{r-s+1} \alpha_i q_{r-i,s-1}
\ee
with $q_{0,0}\equiv1$ and $q_{r,0}\equiv0\,\,(r=1,2,3,\dots)$. Plugging (\ref{X=x:lagrange-inversion}) into the commutator (\ref{X=x:commutator-xp-low-p}) results in
\be
\commij = \ih \sum_{n=0}^\infty \alpha_n \left( \dij + n\frac{\hpi\hpj}{\hp^2} \right) \hp^n \left( \sum_{r=0}^\infty A_r \hp^r\right)^n.
\ee
Further simplification of this expression uses well-known formula for a power series raised to the $n$th power,
\be\label{series-raised-to-power}
\left( \sum_{r=0}^\infty A_r \hp^r\right)^n = \sum_{m=0}^\infty c_m^{(n)} \hp^m,
\ee
where $c_0^{(n)} = 1$ and
\be\label{coefficients-c}
c_m^{(n)} = \frac{1}{m} \sum_{s=1}^m (sn-m+s)A_s c_{m-s}^{(n)} \quad (m\ge1).
\ee
Notice inspection of (\ref{series-raised-to-power}) for $n=0$ immediately reveals $c_m^{(0)}=0$ for any $m\ge1$. Hence, the commutator $\commij$ reads
\be
\commij = \ih \sum_{m=0}^\infty \sum_{n=0}^\infty \alpha_n c_m^{(n)} \left( \dij + n\frac{\hpi\hpj}{\hp^2} \right) \hp^{m+n}.
\ee
This expression can be recast in a more useful form by writing it as a sum on specific powers $\calp$ of $\hp$. For that, we set $\calp=m+n$ so that $m=\calp-n\ge0$ implies $n\le\calp$. The final result is
\be\label{X=x:commutator-xp}
\commij = \ih \sum_{\calp=0}^\infty \sum_{n=0}^\calp \alpha_n c_{\calp-n}^{(n)} \left( \dij + n\frac{\hpi\hpj}{\hp^2} \right) \hp^\calp.
\ee
This is the general expression for the commutator $\commij$ with operators $\hxi=\hXi$ and $\hpi=h(\hrho)\hrhoi$. Beware this commutator can be recast in the deceptively simpler form $F(\hp)\dij + G(\hp)\hpi\hpj$, but the fact is that functions $F(\hp)$ and $G(\hp)$ \textit{cannot} be chosen independently as we note by looking at their series expansion inferred by comparison to (\ref{X=x:commutator-xp}). At last, for future reference, coefficients $c_{\calp-n}^{(n)}$ needed for computations up to $\calp=4$ are listed on Table \ref{table:coeff-c}.

\begin{table}[tb]
\centering
\caption{Coefficients $c_{\calp-n}^{(n)}$ relevant for computation of $\commij$ up to $\oo{\hP^4}$.}
\begin{tabular}{@{\extracolsep{2mm}} c | c c c c c}
\hline\hline
\backslashbox{$\calp$}{$n$} & $0$ & $1$ & $2$ & $3$ & $4$ \TT \BB \\ \hline
$0$ & $1$ & - & - & - & - \TT\BB \\
$1$ & $0$ & $1$ & - & - & - \TT\BB \\
$2$ & $0$ & $-\alpha_1$ & $1$ & - & - \TT\BB \\
$3$ & $0$ & $2\alpha_1^2-\alpha_2$ & $-2\alpha_1$ & $1$ & - \TT\BB \\
$4$ & $0$ & $-5\alpha_1^3+5\alpha_1\alpha_2-\alpha_3$ & $5\alpha_1^2-2\alpha_2$ & $-3\alpha_1$ & $1$ \TT\BB \\ \hline\hline
\end{tabular}
\label{table:coeff-c}
\end{table}

In later sections we compare this commutator to that of other proposals in the literature aiming at having all of them connected to the same general iGUP framework. In particular, this means we will relate each $\alpha_n$ on (\ref{X=x:commutator-xp}) to GUP parameters of each proposal. For instance, computing $\commij$ to $\calo(P^2)$,
\begin{align}\label{X=x:commutator-xp-second-order}
\commij = & \ih\Big[ \dij + \alpha_1 \Big(\hp\dij + \frac{\hpi\hpj}{\hp}\Big) \nonumber\\
& + (\alpha_2 - \alpha_1^2)\hp^2\dij + (2\alpha_2 - \alpha_1^2)\hpi\hpj \Big],
\end{align}
reveals that, except for a slightly different notation, this commutator agrees with Hossenfelder's derivation \cite{hossen2013} and, for the particular case $\alpha_1=-\alpha$ and $\alpha_2=2\alpha^2$, it recovers Ali, Das, and Vagenas' GUP \cite{ali-das-vagenas09,ali-das-vagenas09-rel}.

%%%%%%%%%%%%%%
\subsection{Equivalent commutative iGUP models}
\label{sec:equivalent-models}
%%%%%%%%%%%%%%

Commutative iGUP is obtained setting to zero the sum inside square brackets in (\ref{commutator-xx}). There are basically two routes to implement this condition:
\begin{itemize}
\item[(i)] Setting $\hpi=\hPi$ and $\hxi$ as in (\ref{P=p:operator-x}) but with $f(\hp)$ and $g(\hp)$ satisfying the commutativity condition (\ref{commutativity-condition}). The commutation relation for the position and momentum operators is given by (\ref{P=p:commutator-xp}).
\item [(ii)] Setting $\hxi=\hXi$ and $\hpi(\brho)$ as in (\ref{X=x:operator-P}). It states the obvious fact that any iGUP model with ``standard'' position operator is commutative on position coordinates. The commutator of position and momentum in this case is (\ref{X=x:commutator-xp}).
\end{itemize}
Both approaches lead to equivalent iGUP models as long as they share the same commutator of position and momentum. Comparing (\ref{P=p:commutator-xp}) and (\ref{X=x:commutator-xp}),
we find this happens for
\be\label{alfa-a-1-2}
a_\calp = \sum_{n=0}^{\calp} \alpha_n c_{\calp-n}^{(n)}
\quad \text{and} \quad
b_{\calp-1} = \sum_{n=1}^{\calp+1} n\alpha_n c^{(n)}_{\calp-n+1}.
\ee
Since $c_{\calp-n}^{(n)}$ depends at most on $\alpha_{\calp-n}$ (\cf Table \ref{table:coeff-c}), the first equation above determines $a_\calp$ given $\alpha_\calp$ (or the contrary) order by order starting from $a_0=\alpha_0\equiv1$. The first few $a_n$ in terms of $\alpha_n$, constructed with the help of Table \ref{table:coeff-c}, are
\be
\begin{array}{l} 
a_1 = \alpha_1,
\\[\medskipamount]
a_2 = \alpha_2-\alpha_1^2,
\\[\medskipamount]
a_3 = \alpha_3-3\alpha_2\alpha_1+2\alpha_1^3,
\\[\medskipamount]
a_4 = \alpha_4 -4\alpha_3 \alpha_1 -2\alpha_2^2 +10\alpha_2\alpha_1^2 -5\alpha_1^4.
\end{array}
\ee
Analogously, the second equation on (\ref{alfa-a-1-2}) determines $b_n$ given $\alpha_n$, but amounts to no extra constraint because here $a_n$ and $b_n$ are related by the commutativity condition (\ref{commutativity-condition}) --- \cf comment after (\ref{P=p:commutator-xx-power-series}). The first few $b_n$ in terms of $\alpha_n$ are
\be
\begin{array}{l} 
b_{-1} = \alpha_1,
\\[\medskipamount]
b_0 = 2\alpha_2 -\alpha_1^2,
\\[\medskipamount]
b_1 = 3\alpha_3 -5\alpha_2\alpha_1 +2\alpha_1^3,
\\[\medskipamount]
b_2 = 4\alpha_4 -10\alpha_3\alpha_1 -4\alpha_2^2 +15\alpha_2\alpha_1^2 -5\alpha_1^4.
\end{array}
\ee
In this case, the two routes (i) and (ii) for commutative iGUP are equivalent and the choice of whether working with one or the other is a question of mathematical easiness, but we emphasize that experimental constraints on iGUP parameters $a_n$, $b_n$, and $\alpha_n$ are generally \textit{not} equivalent although easily relatable.

We close this section mentioning two approaches for constructing the representation $\hxi=\hXi$ with $\hpi=h(\hrho)\hrhoi=(\sum\alpha_n\hp^n)\hrhoi$ for any specific commutative iGUP model. One is identifying the functions $f(\hp)$ and $g(\hp)$ by inspection of $\commij$ to extract the parameters $a_n$ and $b_n$ and use them to determine $\alpha_n$ from (\ref{alfa-a-1-2}). This is especially suited for perturbative calculations as it gives $h(\hrho)$ as a power series on $\hp$. The other approach is to consider the one-dimensional reduction $[\hx,\hp]=\ih(f+g\hp^2)$ of (\ref{P=p:commutator-xp}) and the identity $[\hx,\hp(\hrho)]=\ih d\hp/d\hrho$ to solve for $\hrho=\hrho(\hp)$ and use it to identify $\hp/\hrho=h(\hrho)$ under the condition $\hp\to\hrho$ for vanishing iGUP parameters. Whenever the last step gives an exact relation between $\hpi$ and $\hrhoi$, this approach is well-suited for nonperturbative investigations.

%%%%%%%%%%%%%%
\subsection{Three-dimensional extensions of one-dimensional models}
\label{sec:3d-extension-from-1d}
%%%%%%%%%%%%%%

The rise of research on GUP was mostly motivated by the appearance of what has been interpreted as a one-dimensional generalized uncertainty principle in gedanken experiments taking into account gravity in a quantum mechanical context. Not surprisingly, a great deal of attention has been directed to one-dimensional GUP models. This approach revolves around proposing the commutator
\be\label{1-dim-reduction}
[\hx,\hp] = \ih \mathcal{F}
\ee
with a specific function $\mathcal{F}$ of the operator $\hp$. The advantage of this approach is making the whole analysis of the model simpler --- for instance, one obvious choice of operators satisfying this commutator is $\hx=\mathcal{F}\hX$ and $\hp$ with $[\hX,\hp]=\ih$, while another is $\hx$ and $\hp(\hrho)$ with $d\hp/d\hrho=\mathcal{F}(\hp)$ and $[\hx,\hrho]=\ih$.

A reasonable question is what may be understood as the one-dimensional version of a three-dimensional iGUP, for which we see two possibilities: (1) the same component commutator, \eg $[\hx_1,\hp_1]=\ih\mathcal{F}$; or (2) the one-dimensional reduction, $\commij\to[\hx,\hp]=\ih\mathcal{F}$. In general, these two possibilities are associated with \textit{physically different} iGUP models. Exceptions are those based on
\be
\commij=\ih\mathcal{F}\dij,
\ee
which provides a \textit{noncommutative} generalization for any one-dimensional model of the form (\ref{1-dim-reduction}), and for which we remark the choice $\hxi=\hXi$ is \textit{not} allowed even though the one-dimensional model would suggest otherwise. On the other hand, it is then interesting to note this noncommutative model leads to the same one-dimensional \textit{reduction} as any other based on
\be\label{3-dim-extension}
\commij = \ih\Big[ f\dij + \round{\mathcal{F}-f} \frac{\hpi\hpj}{\hp^2} \Big].
\ee
Interestingly enough, (\ref{3-dim-extension}) may even represent a \textit{commutative} model as long as
\be\label{f-comm}
f(\hp) = \frac{\hp}{\int \mathcal{F}^{-1}d\hp} \quad \text{and} \quad f(0)=1,
\ee
where $f$ comes here as the general solution of the nonlinear first order differential equation $\hP\mathcal{F}f' -\mathcal{F}f +f^2 =0$ obtained after imposing the commutativity condition (\ref{commutativity-condition}).

The above simple result exposes the limitations of any approach based on \textit{reduction} to one-dimensional models: conclusions from it cannot be unambiguously traced back to any realistic three-dimensional model. In particular, there are infinitely many three-dimensional noncommutative models (\ref{3-dim-extension}) with the same one-dimensional reduction (\ref{1-dim-reduction}) and at most one commutative model (\ref{f-comm}) sharing this same one-dimensional reduction.

This conclusion is especially important when discussing bounds on different models (see Sec.\ \ref{sec:bounds-igup}). In particular, effectively one-dimensional experiments place bounds on parameters appearing on (\ref{1-dim-reduction}). Setting $\mathcal{F}(\hp)$ as a power series with coefficients $\mathcal{F}_\calp$ with $\mathcal{F}_0=1$, comparison of (\ref{1-dim-reduction}) to the one-dimensional reduction of the commutator (\ref{P=p:commutator-xp-power-series}) for a generally noncommutative iGUP reveals the relation
\be\label{parameters-noncomm}
a_\calp+b_{\calp-2}=\mathcal{F}_\calp
\quad \text{for} \quad
\calp\ge1,
\ee
which expresses that a one-dimensional model characterized by coefficients $\mathcal{F}_\calp$ can be extended to infinitely many noncommutative three-dimensional models characterized by any set of $a_\calp$ and $b_\calp$ satisfying the above relation and that bounds on $\mathcal{F}_\calp$ translate into bounds on the combination $a_\calp+b_{\calp-2}$. On the other hand, if we compare (\ref{1-dim-reduction}) to the one-dimensional reduction of the commutator (\ref{X=x:commutator-xp}) for a commutative iGUP, the extra constrain
\be\label{parameters-comm}
a_\calp+b_{\calp-2}=\sum_{n=1}^\calp(1+n)\alpha_n c_{\calp-n}^{(n)} = \mathcal{F}_\calp
\quad \text{for} \quad
\calp\ge1
\ee
is enforced. Here $a_\calp$ and $b_\calp$ are not independent parameters because of (\ref{alfa-a-1-2}). This extra condition enforces that there is at most one three-dimensional commutative extension to (\ref{1-dim-reduction}), which can be characterized by the set of $\alpha_n$. In this sense, bounds on $\mathcal{F}_\calp$ place bounds on combinations of $\alpha_n$ as in the above relation.

%%%%%%%%%%%%%%
\section{Connection to specific GUP proposals}
\label{sec:applications}
%%%%%%%%%%%%%%

In this section we discuss GUP models proposed in the literature in the context of the iGUP framework devised on the previous section. For simplicity, we restrict ourselves to consider a few representative proposals only. We concentrate on verifying the consistency of the approach and providing alternative, general routes to some results on model construction. Relation of the models' parameters to iGUP parameters $a_n$, $b_n$, and $\alpha_n$ will be emphasized as necessary according to further convenience. As a necessity, we adapt notation used by other authors to that adopted here.

%%%%%%%%%%%%%%
\subsection{Kempf-Mangano-Mann's one-dimensional GUP}
\label{sec:kmm-1d}
%%%%%%%%%%%%%%

The arguably most popular one-dimensional GUP model was proposed by Kempf, Mangano, and Mann (KMM) \cite{kmm95}. In our notation, it is given by the commutator
\be\label{applications:kmm-1d}
[\hx,\hp] = \ih (1 +\beta\hp^2)
\ee
and predicts a minimum uncertainty of $\hbar\sqrt{\beta}$ on position measurements. For their analysis, KMM proposed the representation $\hx=(1+\beta\hp^2)\hX$ with $\hp=\hP$. Latter, an alternative representation also satisfying (\ref{applications:kmm-1d}) was proposed by Pedram \cite{pedram12-prd}, namely $\hx = \hX$ and $\hp = \tan (\sqrt{\beta}\hrho)/\sqrt{\beta}$.

Identification of specific iGUP parameters $a_n$ or $b_n$ for this model is meaningless because there are infinitely many noncommutative three-dimensional models reducing to this one-dimension model and the only reasonable identification is $a_2+b_0=\beta$ (\cf Sec.\ \ref{sec:3d-extension-from-1d}). On the other hand, there is only one commutative three-dimensional extension and it can be characterized by the set of $\alpha_n$, identified here as the coefficients of the MacLaurin series expansion of $\tan (\sqrt{\beta}\hrho)/\sqrt{\beta}\hrho$ for Pedram's choice of operators. In particular, the first few iGUP parameters $\alpha_n$ of KMM's model are $\alpha_0=1$, $\alpha_1=0$, $\alpha_2=\frac{1}{3}\beta$, $\alpha_3=0$ and $\alpha_4=\frac{2}{15}\beta^2$.

%%%%%%%%%%%%%%
\subsection{Kempf-Mangano-Mann's three-dimensional GUP}
\label{sec:kmm-3d}
%%%%%%%%%%%%%%

KMM also proposed an immediate three-dimensional extension of (\ref{applications:kmm-1d}) given by the commutators
\be\label{applications:kmm-3d-noncomm}
\commij = \ih\dij (1+\beta\hp^2),
\ee
\be
\commxx = -2\ih\beta(\hxi\hpj -\hxj\hpi),
\ee
with the choice $\hxi = (1+\beta\hp^2)\hXi$ and $\hpi = \hPi$ \cite{kmm95}. This version predicts a minimum position uncertainty of $\hbar\sqrt{3\beta}$ instead \cite{kempf97}, thus the GUP parameter of both three and one-dimensional models would be better denoted by $\beta_\text{3D}$ and $\beta_\text{1D}$, respectively, where $\beta_\text{3D}=3\beta_\text{1D}$. Except for that, the only nontrivial iGUP parameter for this model is $a_2=\beta$.

Since this three-dimensional extension is noncommutative on position coordinates, an operator representation based on $\hxi=\hXi$ analogous to that proposed by Pedram for the one-dimensional model cannot be simply extended for the three-dimensional case. On the other hand, if one sticks with the one-dimensional commutator (\ref{applications:kmm-1d}), a \textit{different} three-dimensional extension of the form $\commij=\ih[f\dij+(1+\beta\hp^2-f)\hpi\hpj/\hp^2]$ can be found for \textit{commuting} position operators as long as $f$ is given by (\ref{f-comm}). In this case, we find
\begin{eqnarray}\label{applications-kmm-3d-comm}
\commij & = & \ih \Bigg[ \frac{ \sqrt{\beta}\hp }{ \tan^{-1}(\sqrt{\beta}\hp) }\dij \nonumber\\
& & + \Bigg( 1+\beta\hp^2 - \frac{ \sqrt{\beta}\hp }{ \tan^{-1}(\sqrt{\beta}\hp) } \Bigg) \frac{\hpi\hpj}{\hp^2} \Bigg],
\end{eqnarray}
\be
\commxx\equiv0
\ee
is the only commutative three-dimensional extension of (\ref{applications:kmm-1d}). To investigate physical consequences of this model, one choice for position and momentum operators is $\hpi=\hPi$ with $\hxi$ given by (\ref{P=p:operator-x}) after comparing (\ref{applications-kmm-3d-comm}) to (\ref{P=p:commutator-xp}), but another, much more convenient choice is
\be
\hxi=\hXi
\quad \text{and} \quad
\hpi = \hrhoi \frac{\tan(\sqrt{\beta}\hrho)}{\sqrt{\beta}\hrho}
\ee
which extends Pedram's approach for the three-dimensional case indeed, a result recently obtained on \cite{chung19} by completely different means. The equivalence of the two representations is straightforwardly, although tediously as well, verified on the grounds discussed in Sec.\ \ref{sec:equivalent-models}.

%%%%%%%%%%%%%%
\subsection{Kempf's GUP}
\label{sec:kempf}
%%%%%%%%%%%%%%

Kempf proposed the three-dimensional commutator
\be\label{application:kempf}
\commij = \ih\Big[ (1+\beta\hp^2)\dij + \beta'\hpi\hpj \Big],
\ee
understood as the lowest order isotropic correction (on even powers of the momentum) to the canonical commutation relation, which predicts $\hbar\sqrt{3\beta+\beta'}$ for the minimal uncertainty on position \cite{kempf97}. Considering the iGUP framework, this model is obtained setting the iGUP parameters to $a_2=\beta$ and $b_0=\beta'$, with all the others vanishing. For the position operator (\ref{P=p:operator-x}),
\be
\hxi = (1+\beta\hp^2)\hXi + \beta'\hpi(\pdotx) +\ih\gamma(\hp)\hpi,
\ee
the function $\gamma(\hp)$ is arbitrary in the sense discussed in Sec.\ \ref{sec:P=p}, where the choice $\gamma=f'/2\hP+2g = \beta+2\beta'$ based on (\ref{P=p:scalar-product}) reproduces the representation chosen in \cite{kempf97}. The commutator between position operators (\ref{P=p:commutator-xx}) reads
\be
\commxx = \ih \frac{(2\beta-\beta') + (2\beta+\beta')\beta\hp^2}{1+\beta\hp^2} (\hpi\hxj - \hpj\hxi)
\ee
and indicates that the particular case of $\beta'=2\beta$ is \textit{approximately} commutative on the position operators, $\commxx=0$ to $\oo{\beta}$. Only to this order, another acceptable choice of operators, \eg obtained after solving (\ref{X=x:operator-P}) or (\ref{alfa-a-1-2}), is $\hxi=\hXi$ and $\hpi=\hrhoi(1+\beta\hrho^2)$, which allows identification of the relevant iGUP parameter $\alpha_2=\beta$.

%%%%%%%%%%%%%%
\subsection{Kempf-Mangano commutative GUP}
%%%%%%%%%%%%%%

Kempf and Mangano studied to great detail the model based on \cite{kempf-mangano97}
\be
\commij = \ih ( f\dij + 2\beta\hpi\hpj ).
\ee
Setting the choice $\hpi=\hPi$, they required the model to be commutative on components of the position operator. In the iGUP framework, this is enforced by the commutativity condition (\ref{commutativity-condition}), which guarantees components of the position operator $\hxi = f\hXi + 2\beta\hpi(\pdotx) +\ih\gamma\hpi$ commute with each other as long as
\be
f(\hp) = \frac{1}{2}\round{1+ \sqrt{1+4\beta\hp^2} } = \frac{2\beta\hp^2}{\sqrt{1+4\beta\hp^2} -1},
\ee
as also found by Kempf and Mangano. The minimal uncertainty $2.29\hbar\sqrt{\beta}$ they find for position measurements indicates position space representation of the wavefunction is not achievable, but because the model is commutative on $\hxi$, the alternative choice of operators $\hxi=\hXi$ with $\hpi=\hrhoi/(1-\beta\hrho^2)$ is a viable option nevertheless.

%%%%%%%%%%%%%%
\subsection{Ali-Das-Vagenas's GUP}
\label{sec:ali-das-vagenas}
%%%%%%%%%%%%%%

Ali, Das, and Vagenas (ADV) proposed the commutator
\begin{align}\label{application:ali-das-vagenas-comm-xp}
\commij = \ih\Big[ \dij - \alpha \Big(\hp & \dij + \frac{\hpi\hpj}{\hp}\Big) \nonumber\\
& + \alpha^2(\hp^2\dij +3\hpi\hpj) \Big],
\end{align}
predicting not only a minimum position uncertainty of $\hbar\alpha$ but also a maximum momentum uncertainty of $\alpha^{-1}$ \cite{ali-das-vagenas09}. The model is also commutative on position operators to $\oo{\alpha^2}$; thus, within our framework, the above commutator may be expressed as (\ref{P=p:commutator-xp}) or equivalently as (\ref{X=x:commutator-xp}). From (\ref{P=p:commutator-xp}), we identify
\be
f(\hp) = 1-\alpha\hp +\alpha^2\hp^2
\quad \text{and} \quad
g(\hp) = - \frac{\alpha}{\hp} + 3\alpha^2.
\ee
Hence, $a_1=-\alpha$, $a_2=\alpha^2$, $b_{-1}=-\alpha$ and $b_0=3\alpha^2$ and the representation $\hpi=\hPi$ is accomplished along with
\be
\hxi = (1-\alpha\hp +\alpha^2\hp^2)\hXi - \Big( \frac{\alpha}{\hp} - 3\alpha^2 \Big) \hpi(\pdotx) +\ih\gamma\hpi,
\ee
which satisfies the commutativity condition (\ref{commutativity-condition}) to $\oo{\alpha^2}$ indeed. For an alternative representation, from (\ref{X=x:commutator-xp}) we notice ADV's commutator exactly matches the $\oo{\hP^2}$ commutator (\ref{X=x:commutator-xp-second-order}) for $\alpha_1=-\alpha$ and $\alpha_2=2\alpha^2$; thus, the representation $\hxi=\hXi$ comes along with
\be
\hpi = (1 -\alpha\hrho +2\alpha^2\hrho^2)\hrhoi,
\ee
which agrees with \cite{ali-das-vagenas09}.

%%%%%%%%%%%%%%
\subsection{Pedram's GUP}
%%%%%%%%%%%%%%

Pedram proposed a specific nonperpertubative GUP characterized by the one dimensional commutator \cite{pedram12-plb2}
\be\label{application:pedram-1d}
[\hx,\hp] = \frac{\ih}{1-\beta\hp^2}
\ee
and proceeded proposing a natural generalization for higher dimensions \cite{pedram12-plb3},
\be\label{application:pedram-3d}
\commij = \frac{\ih\dij}{1-\beta\hp^2}.
\ee
Besides predicting a nonvanishing minimum position uncertainty, an attractive feature of this model coming from its nonperturbative nature is the upper bound on the physical momentum, $p<1/\sqrt{\beta}$. This three-dimensional model is generated in the iGUP framework setting $f(\hp)=1/(1-\beta\hp^2)$, $g(\hp)=0$, and $h(\hP)=1$; thus, we notice the choice
\be
\hxi = \frac{1}{1-\beta\hp^2}\hXi + \ih\gamma(\hp)\hpi
\quad \text{and} \quad
\hpi=\hPi
\ee
and the consequent commutator
\be
\commxx = \frac{2\ih\beta}{(1-\beta\hp^2)^2} (\hpi\hxj-\hpj\hxi)
\ee
both agree with those found on \cite{pedram12-plb3}.

Following the discussion in Sec.\ \ref{sec:3d-extension-from-1d}, the commutative three-dimensional extension of (\ref{application:pedram-1d}) is
\be
\commij = \ih\Bigg[ \frac{\dij}{1-\frac{1}{3}\beta\hp^2} +\frac{\frac{2}{3}\beta}{(1-\frac{1}{3}\beta\hp^2)(1-\beta\hp^2)}\hpi\hpj \Bigg],
\ee
in agreement with Ref.\ \cite{shababi-chung17}. Comparison to (\ref{P=p:commutator-xp}) immediately reveals the form of $\hxi(\bX,\bP)$ for the choice $\hpi=\hPi$. On the other hand, another equivalent choice sets $\hxi=\hXi$ and $\hpi(\brho)$. As an illustration, it can be derived from the first relation on (\ref{alfa-a-1-2}), translated here to
\be
\sum_{n=1}^\calp \alpha_n c_{\calp-n}^{(n)} = \left\{
\begin{array}{cl}
  (\frac{1}{3}\beta)^{\calp/2} & \text{for even $\calp$,} \\[\medskipamount]
 0 & \text{for odd $\calp$,}
\end{array}
\right.
\ee
after solving order by order for $\alpha_n$ and using it to write
\be\label{applications:pedram-perturbative}
\hpi = \round{1 +\frac{1}{3}\beta\hp^2 +\frac{1}{3}\beta^2\hp^4 +\frac{4}{9}\beta^3\hp^6 +\cdots} \hrhoi,
\ee
which is the series expansion also found on \cite{pedram12-plb3,chung-hass-int-j}. The other approach for finding $\hpi$ discussed at the end of Sec.\ \ref{sec:equivalent-models} leads to an equivalent, closed expression
\be\label{applications:pedram-nonperturbative}
\hpi = \frac{1 -i\sqrt{3} +(-2\beta)^{1/3}( 3\hp +\sqrt{9\hp^2-4/\beta} )^{2/3} }{ (2\beta)^{2/3} ( 3\hp +\sqrt{9\hp^2-4/\beta} )^{1/3} } \frac{\hrhoi}{\hrho},
\ee
also  found on Ref.\ \cite{pedram12-plb3}. Although experimental bounds on model parameters are usually set using perturbative calculations, and for that (\ref{applications:pedram-perturbative}) is very suited, theoretical insights are usually obscured in this approach. For instance, only with the closed form (\ref{applications:pedram-nonperturbative}) we notice the generator of translations $\hrhoi$ is also bounded from above, $\rho\le 2/(3\sqrt{\beta})$.

%%%%%%%%%%%%%%
\subsection{Chung-Hassanabadi's GUP}
%%%%%%%%%%%%%%

More recently, Chung and Hassanabadi proposed the one-dimensional commutator 
\be\label{chung-hass-1d}
[\hx,\hp] = \frac{\ih}{(1-\beta\hp)^N},
\ee
where $\hp$ stands for $|\hp|$ \cite{chung-hass2019}. Similar to Pedram's model, besides predicting a nonvanishing minimum position uncertainty, this nonperturbative model also predicts an upper bound on momentum eigenvalues, namely $|p|\ge 1/\beta$ --- but notice this $\beta$ is not directly related to that of KMM's or Pedram's GUP since here $[\beta]=[p]^{-1}$, being better identified with the $\alpha$ parameter of models inspired by ADV's. Following the discussion in Sec.\ \ref{sec:3d-extension-from-1d}, the commutative three-dimensional extension of this proposal is found to be
\begin{align}
\commij = \ih\Bigg[ & \frac{\beta(1+N)\hp}{ 1-(1-\beta\hp)^{1+N} }\dij \nonumber\\
& + \frac{(1-\beta\hp)^{-N} - N\beta\hp-1}{1-(1-\beta\hp)^{1+N}} \frac{\hpi\hpj}{\hp^2} \Bigg],
\end{align}
which, as far as we know, is new to the literature. At last, proceeding analogously to what we described by the end of the previous section, an operator representation based on $\hxi=\hXi$ can be derived, resulting in
\begin{align}
\hpi & = \hrhoi \squ{1 +\frac{N}{2}\beta\hrho +\frac{N(1+2N)}{6}(\beta\hrho)^2 +\dots } \nonumber\\
& = \hrhoi \sum_{n=0}^\infty \begin{pmatrix} \frac{1}{1+N}\\n \end{pmatrix} (1+N)^n (\beta\hrho)^n \nonumber\\
& = \hrhoi \frac{1-\squ{1 -(1-N) \beta\hrho}^{1/(1-N)} }{\beta\hrho},
\end{align}
which agrees with the one-dimensional version on \cite{chung-hass2019}.

%%%%%%%%%%%%%%
\subsection{Petruzziello's GUP}
%%%%%%%%%%%%%%

As a last illustration of the iGUP framework, we consider the very recent one-dimensional model proposed by Petruzziello \cite{petruzziello2020} (see also \cite{maggiore2021}),
\be\label{application:petruzziello}
[\hx,\hp] = \ih \sqrt{1-2|\beta|\hp^2} = \ih\round{ 1 - |\beta|\hp^2 - \cdots },
\ee
which directly relates to a model with $\beta<0$ briefly discussed by Kempf in 1997 \cite{kempf-negative}. This model has a curious feature: the sign of $\beta$ is opposite to that of previous models. Besides featuring a finite maximum value for momentum and momentum uncertainty while allowing for arbitrarily small position uncertainty, such sign choice may provide a different and interesting route for investigation of quantum gravity phenomena as it predicts a classical behavior at high momentum --- in particular, this situation emerges naturally for the lattice spacetime model discussed on \cite{scardigli2010}.

The study of physical consequences of Petruzziello's proposal on Ref.\ \cite{petruzziello2020} was mainly devoted to black hole thermodynamics so there is a lot of room for further investigation. In particular, with the results in Sec.\ \ref{sec:3d-extension-from-1d} we can extend this one-dimensional proposal to the three-dimensional commutative iGUP characterized by
\begin{align}
\commij = \,& \ih\Bigg[ \frac{\sqrt{2\beta}\hp}{\sin^{-1}(\sqrt{2\beta}\hp)}\dij \nonumber\\
& + \round{ \sqrt{1-2\beta\hp^2} - \frac{\sqrt{2\beta}\hp}{\sin^{-1}(\sqrt{2\beta}\hp)}} \frac{\hpi\hpj}{\hp^2} \Bigg],
\end{align}
satisfied, for instance, setting $\hpi=\hPi$ and
\begin{align}
\hxi = \,& \frac{\sqrt{2\beta}\hp}{\sin^{-1}(\sqrt{2\beta}\hp)}\hXi + \Bigg( \sqrt{1-2\beta\hp^2} \nonumber\\
& - \frac{\sqrt{2\beta}\hp}{\sin^{-1}(\sqrt{2\beta}\hp)} \Bigg)\frac{\hpi}{\hp^2}(\pdotx) +\ih\gamma(\hp)\hpi
\end{align}
or the far more viable choice setting $\hxi=\hXi$ and
\be
\hpi = \hrhoi \frac{\sin(\sqrt{2\beta}\hrho)}{\sqrt{2\beta}\hrho}.
\ee
For a perturbative approach up to $\oo{\beta}$, the commutator is simplified to $\commij=\ih[(1-\frac{1}{3}\beta\hp^2)\dij -\frac{2}{3}\beta\hpi\hpj]$; the first representation to $\hxi=(1-\frac{1}{3}\beta\hp^2)\hXi -\frac{2}{3}\beta\hpi(\pdotx) +\ih\gamma(\hp)\hpi$ and $\hpi=\hPi$; and the second representation to $\hxi=\hXi$ and $\hpi=(1-\frac{1}{3}\beta\hrho^2)\hrhoi$.

%%%%%%%%%%%%%%
\section{Translating bounds on iGUP parameters}
\label{sec:bounds-igup}
%%%%%%%%%%%%%%

To discuss bounds on iGUP models one may prefer to start separating them into two distinct classes:
\begin{itemize}
\item[(i)] \textit{Noncommutative models.} These contain effectively two families of parameters that, adopting a different notation, are represented by $\breve{a}_n$ and $\breve{b}_n$ considering the operator representation $\hxi(\bX,\bp)$ and $\hpi$ with $[\hXi,\hpj]=\ih\dij$ of Sec.\ \ref{sec:P=p}. For any model, there are other equivalent, ``mixed'' representations with $[\hXi,\hpj]\neq\ih\dij$ instead, for which we adopt the notation $\breve{a}^\times_n$, $\breve{b}^\times_n$, and $\breve{\alpha}^\times_n$, noticing observable combinations of these reduce them to effectively two families only. For simplicity, all bounds we report here will be translated into bounds on $\breve{a}_n$ and $\breve{b}_n$.
\item[(ii)] \textit{Commutative models.} For these, there is effectively only one family of parameters, $\alpha_n\sim a_n\sim b_n$, because in this case these parameters relate to each other due to the equivalence of the representations $\hxi(\bX,\bp)$ and $\hpi$ with $[\hXi,\hpj]=\ih\dij$ (Sec.\ \ref{sec:P=p}); $\hxi$ and $\hpi(\brho)$ with $[\hxi,\hrhoj]=\ih\dij$ (Sec.\ \ref{sec:X=x}); and mixed ones with $\hxi(\bX,\bP)$ and $\hpi(\bP)$ where $[\hXi,\hPj]=\ih\dij$ --- parameters of the first two representations, for instance, are related by (\ref{alfa-a-1-2}). For simplicity, bounds on these models will be translated into bounds on $\alpha_n$.
\end{itemize}
For any truly three-dimensional experiment, both classes may be well distinguished. On the other hand, effectively one-dimensional experiments cannot distinguish between the two. To any attainable power $\calp\ge1$ on $\hP$, one-dimensional experiments place simultaneous bounds on the combination $\breve{a}_\calp+\breve{b}_{\calp-2}$ of noncommutative models (\ref{parameters-noncomm}) and on the combination $\sum (1+n)\alpha_n c_{\calp-n}^{(n)}$ of commutative models (\ref{parameters-comm}).

Most experimental bounds available on the literature, which we discuss next, are placed on KMM's one-dimensional model (Sec.\ \ref{sec:kmm-1d}), and the three-dimensional models of Kempf (Sec.\ \ref{sec:kempf}) and Ali-Das-Vagena (Sec.\ \ref{sec:ali-das-vagenas}). The relevant parameter of both KMM's and Kempf's (approximately commutative) model is $\beta$, and bounds are usually expressed in terms of the dimensionless parameter $\beta_0$ by $\beta=\beta_0\ell_P^2/\hbar^2$; while that of ADV's model is $\alpha$, expressed in terms of the also dimensionless $\alpha_0$ by $\alpha=\alpha_0\ell_P/\hbar$ --- beware this $\alpha_0$ has no relation to iGUP's $\alpha_0\equiv1$ (Sec.\ \ref{sec:isotropic-operators}). To keep the notation simple, bounds on iGUP parameters will always be expressed in units of $\ell_P/\hbar$. In what follows we translate bounds on these models into bounds on iGUP parameters after briefly explaining their origin. A summary is given at the end of this section.

\subsection{Bounds on KMM's one-dimensional model from nongravitational tests}

To the best of our knowledge, current \textit{direct} bounds on KMM's one-dimensional model (Sec.\ \ref{sec:kmm-1d}) all come from experiments dealing with mechanical oscillators --- in particular, composite bodies. An implicit \textit{assumption} is that they are subjected to the same deformed commutator a point particle would be, with same GUP coefficient $\beta$, but this hypothesis may have significant impact on the resulting bounds on $\beta$ \cite{composite}, which should therefore be interpreted with care.

Back in 2013, Marin \textit{et al.} reported on an experiment consisting on the measurement of the energy $E_\text{exp}$ of the first longitudinal mode of the sub-milikelvin cooled ton-scale bar used on the gravitational wave detector AURIGA \cite{marin}. The theoretical prediction for the lowest energy $E_\text{min}$ was obtained after saturating the uncertainty principle $\Delta x\cdot\Delta p\ge\frac{\hbar}{2}(1+\beta \la\hp^2\ra)$ considering the bar oscillated as a simple harmonic oscillator described by operators $\hx$ and $\hp$ satisfying $[\hx,\hp]=\ih(1+\beta\hp^2)$ instead of the conventional canonical commutation relation. Direct comparison $E_\text{min}<E_\text{exp}$ then resulted in the upper limit $\beta_0<3\times10^{33}$.

Stronger bounds from a dedicated experiment were reported two years later by Bawaj \textit{et al.} \cite{bawaj2005}. They tracked the time-evolution of high-Q micro and nano mechanical oscillators with masses ranging from $2\times10^{-11}$ to $3.3\times10^{-5}$ kg looking for two predictions of the model: a residual frequency fluctuation ($\Delta\omega$) to the frequency dependence on the amplitude, and the appearance of a third harmonic on the oscillation modes. For the lighter oscillator came stronger bounds, $\beta_0<3\times10^7$ from the residual frequency $\Delta\omega$ and $2\times10^{11}$ from the third harmonic analysis; for the heaviest oscillator weaker bounds were derived, $\beta_0<2\times10^{19}$ and $1\times10^{26}$ from analysis of $\Delta\omega$ and the third harmonic, respectively. 

Recently, Bushev \textit{et al.} used a 0.3 kg ultra-high-Q sapphire split-bar mechanical resonator to search for the predicted amplitude dependence of the frequency, further improving the upper limit on $\beta_0$ to $5.2 \times 10^6$ \cite{bushev}.

As discussed in Sec.\ \ref{sec:kmm-1d}, the model these experiments investigated correspond to the one-dimensional reduction of any noncommutative iGUP model with $\breve{a}_2+\breve{b}_0=\beta\equiv\beta_0\ell^2_P/\hbar^2$ or the commutative model with $\alpha_n$ given by the coefficients of the Mclaurin series expansion of $\tan (\sqrt{\beta}\hrho)/\sqrt{\beta}\hrho$, where the first nontrivial is $\alpha_2=\frac{1}{3}\beta$. Thus, the above experiments place upper limits on these coefficients. In particular, the most stringent bound reported sets $\breve{a}_2+\breve{b}_0 < 5.2\times10^6 \ell_P^2/\hbar^2$ and $\alpha_2 < 1.7\times10^6 \ell_P^2/\hbar^2$.

\subsection{Bounds on KMM's one-dimensional model from gravitational tests}
\label{sec:bound-grav}

Implementation of a generalized uncertainty principle on the study of gravitational phenomena amounts for ongoing discussion \cite{scardigli-casadio}. In particular, a common approach is to derive corrections to classical Newtonian mechanics by extending the quantum commutator $[\hx,\hp]=\ih\mathcal{F}(\hp)$ into modifications on the Poisson brackets, namely  $\{x,p\}=\mathcal{F}(p)$ \cite{pedram12-plb3,gravA1,gravA2}; and similarly for relativistic classical mechanics \cite{gravB1,gravB2,gravB3}. One critique on this approach is that the limit of vanishing GUP parameters recovers only Newtonian mechanics but not general relativity. It is, therefore, implicitly assumed that GUP-corrections and general relativity coexist independently, but it could be argued instead that general relativity itself should receive GUP-corrections. Besides leading to a violation of the equivalence principle (\cf Appendix A on \cite{scardigli-casadio}), as a consequence of this approach the very accurate predictions of general relativity for gravitational phenomena leave basically no room for extra contributions from GUP; for instance, measurements of Mercury's perihelion precession sets the tremendously small bound $\beta_0<10^{-66}$ on KMM's model \cite{gravA1}, which may be questionable on the grounds just discussed (\cf \cite{casadio-scardigli-2020} for a detailed analysis). 

To avoid these issues, Scardigli and Casadio proposed a different approach \cite{scardigli-casadio}. They studied the effects of KMM's uncertainty principle $\Delta x \cdot \Delta p \ge \frac{\hbar}{2} (1 + \beta \la p^2 \ra)$ on the Hawking radiation process and related the corrections to deformations on the Schwarzschild metric. This allowed for direct implementation of GUP corrections into general relativity, particularly preserving the equivalence principle and the standard geodesic equation, without touching the basic structures of classical Newtonian mechanics. Making use of the deformed Schwarzschild metric and precision measurements of gravitational phenomena, they derived several upper limits to $\beta$ (in units of $\ell_P^2/\hbar^2$): $1.3\times10^{78}$ from light deflection by the Sun; $7.5\times10^{71}$ from the precession of Mercury's perihelion; and $5\times10^{70}$ from the binary pulsar PRS B 1913+16 data. Despite not as stringent as the bounds from quantum mechanical systems, these are gravitational bounds derived directly from a generalized uncertainty principle without evoking a particular commutation relation nor requiring a violation of the equivalence principle.

The bounds derived by Scardigli and Casadio apply, based on previous discussions, to iGUP parameter combination $\breve{a}_2+\breve{b}_0$, while for $\alpha_2$ they read $4.3\times10^{77}$ (light deflection by the Sun), $2.5\times10^{71}$ (perihelion precession of Mercury), and $2\times10^{70}$ (pulsar PRS B 1913+16 data).

Another approach for constraining $\beta$, proposed later by Lambiase and Scardigli \cite{gup-sme}, compares GUP-corrections to the Hawking temperature to those derived assuming Lorentz symmetry violations as parametrized by the Standard Model Extension (SME) \cite{sme1,sme2,sme3}. The available tight bounds on SME coefficients (\cf \cite{datatable}) makes a comparison to their isotropic part a source of similarly tight bounds models like iGUP where Lorentz symmetry is assumed as an exact property of flat spacetime. In particular, they use bounds on SME coefficients from torsion pendulum experiments to constrain the parameter $\beta$ at the level of $10^{51} \ell_P^2/\hbar^2$. In the iGUP framework, this corresponds to constraints on $\breve{a}_2+\breve{b}_0$ and $\alpha_2$ of the same order of magnitude.

\subsection{Bounds on Kempf's noncommutative model}

The hydrogen atom is one of the simplest quantum mechanical systems, allows for realistic theoretical modeling and its energy spectrum has been measured to great precision. In the context of GUP scenarios, it is especially suited for testing three-dimensional iGUP models. In particular, Kempf's model (Sec.\ \ref{sec:kempf}) had received detailed investigation in the past years \cite{brau,ak-yao,benczik,stetsko-tkachuk,stetsko,bouaziz}. Of such, Stetsko \cite{stetsko} considered the hydrogen Hamiltonian to $\oo{\beta,\beta'}$ in the ``mixed'' representation $\hxi = \hXi + \frac{1}{4}(2\beta-\beta')(\hXi\hP^2+\hP^2\hXi)$ and $\hpi=\hPi(1+\frac{1}{2}\beta'\hP^2)$ to derive corrections to any $s$th energy level. Experimental data from precision hydrogen spectroscopy was used for the Lamb shift on $s$ levels and a resulting upper limit $\beta+\beta'<10^{37}\ell_P^2/\hbar^2$ was obtained. Although Stetsko performs a detailed analysis using real experimental data, due to intricacies related to using the Lamb shift measurements for deriving bounds on $\beta+\beta'$,
he puts emphasis on the order of magnitude of the bound instead of on actual numbers (\cf Sec.\ III in \cite{stetsko}).

In the context of the iGUP framework, the mixed representation chosen by Stetsko can be rewritten as $\hxi = (1 + \breve{a}^\times_2\hP^2)\hXi + \ih\breve{\gamma}^\times_0\hPi$ and $\hpi=(1+\breve{\alpha}^\times_2\hP^2)\hPi$ with iGUP parameters $\breve{a}^\times_2=\breve{\gamma}^\times_0 = \frac{1}{2}(2\beta-\beta')$ and $\breve{\alpha}^\times_2=\frac{1}{2}\beta'$. Therefore, the bound reported on \cite{stetsko} translates into $\breve{a}^\times_2 + 3\breve{\alpha}^\times_2 < 10^{37} \ell_P^2/\hbar^2$, and there is no bound on $\breve{\gamma}^\times_0$ as expected because it is an unobservable parameter (\cf Sec.\ \ref{sec:P=p}). On the other hand, an equivalent representation is obtained setting iGUP parameters to $\breve{a}_2=\beta$ and $\breve{b}_0=\beta'$ (\cf Sec.\ \ref{sec:kempf}). These parameters are, therefore, bounded by $\breve{a}_2+\breve{b}_0<10^{37}\ell_P^2/\hbar^2$.

Since Kempf's model for the particular case $\beta'=2\beta$ is commutative to order $\oo{\beta}$ on position operators, the above upper limit is valid for the corresponding three-dimensional commutative iGUP model with $\alpha_2=\beta$ as well (\cf Sec.\ \ref{sec:kempf}); thus, $\alpha_2<10^{37}\ell_P^2/\hbar^2$.

\subsection{Bounds on Kempf's approximately commutative model}

For any commutative iGUP model, position and momentum operators can be chosen as $\hxi$ and $\hpi(\brho)$ with $[\hxi,\hrhoj]=\ih\dij$. Quantum gravity effects associated with a modified canonical commutation relation may then be understood as universal as they affect the kinetic part of the Hamiltonian of any system in consideration. In this context, Das and Vagenas set $\hpi=\hrhoi(1+\beta\hrho^2)$ for Kempf's $\oo{\beta}$-commutative model (Sec.\ \ref{sec:kempf}) to derive corrections to the Landau levels of an electron in a magnetic field, then estimating $\beta_0 < 10^{50}$ \cite{das-vagenas08}. Along the same lines, Quesne and Tkachuk considered corrections to the hydrogen atom Hamiltonian but with the extra hypothesis that each type of particle has its own characteristic parameter $\beta_0$ and used precision measurements of the 1S-2S transition to place $\beta_0 <2.23\times10^{34}$ for the electron \cite{quesne-tkachuk10}. As discussed in Sec.\ \ref{sec:kempf}, these bounds (in units of $\ell_P^2/\hbar^2$) are directly set on the iGUP parameter $\alpha_2$.

Effectively one-dimensional systems were also considered to place bounds on this model. In this case, the relevant commutator between position and momentum operators reduces to $[\hx,\hp]=\ih(1+3\beta\hp^2)$, corresponding to three-dimensional noncommutative iGUP models with $\breve{a}_2+\breve{b}_0=3\beta$ or commutative ones with $\alpha_2=\beta$. These are the iGUP parameters to be bounded by Das and Vagenas' estimate $\beta_0<10^{21}$ when considering the effects of this commutator on a scanning tunneling microscope \cite{das-vagenas08}; and also Gao and Zhan's bound $\beta_0<1.3\times10^{39}$ coming from precision measurements of the ratio $\hbar/m_\text{Rb}$ done on \textsuperscript{87}Rb cold atom recoil experiments \cite{gao-zhan}.

In the same spirit of Scardigli and Casadio's approach \cite{gup-sme} mentioned at the end of Sec.\ \ref{sec:bound-grav}, Gomes \cite{nigup} also uses the tight bounds on SME coefficients to constrain GUP models. In particular, it is proposed an extension of Kempf's approximately commutative model to anisotropic spaces. The resulting Hamiltonian is found to be indistinguishable from that of the non-relativistic SME Hamiltonian for fermions, allowing the direct transfer of bounds on SME coefficients from searches of annual variations of the hydrogen 1S-2S energy difference to (non isotropic) GUP parameters. Restricting to the isotropic combination of such parameters allows deriving $\beta < 10^{30}\ell_P^2/\hbar^2$, which directly translates into a bound on the iGUP parameters $\alpha_2$.

\subsection{Bounds on ADV's approximately commutative model}

The approach mentioned in the previous section was also very similarly employed to bound Ali-Das-Vagenas' model, discussed in Sec.\ \ref{sec:ali-das-vagenas}. This model is approximately commutative to $\oo{\alpha^2}$ and allows for the representation $\hxi$ and $\hpi=(1-\alpha\hrho+2\alpha^2\hrho^2)\hrhoi$.
Ali, Das, and Vagenas used this to estimate $\alpha<10^{21}\ell_P/\hbar$ investigating the Landau levels of an electron in a magnetic field. They also considered corrections to the hydrogen atom Hamiltonian, using the accuracy of precision measurements of the Lamb shift to estimate $\alpha<10^{10}\ell_P/\hbar$  \cite{ali-das-vagenas11}. These bounds are placed on iGUP parameters under the identification $\alpha_1=-\alpha$ and $\alpha_2=2\alpha^2$.

ADV's model was also considered in the context of effectively one-dimensional systems. The relevant commutator is $[\hx,\hp]=\ih(1-2\alpha\hp+4\alpha^2\hp^2)$, which is derived from three-dimensional noncommutative iGUP models with $\breve{a}_1+\breve{b}_{-1}=-2\alpha$ and $\breve{a}_2+\breve{b}_{0}=4\alpha^2$ or commutative models with $\alpha_1=-\alpha$ and $\alpha_2=2\alpha^2$. These are the iGUP parameters bounded by the estimate  $\alpha < 10^{17}\ell_P/\hbar$ derived by Ali, Das, and Vagenas for the corrections to the ground state energy of the charmonium, represented as a simple harmonic oscillator \cite{ali-das-vagenas11}; and by $\alpha<2.4\times10^{14}\ell_P/\hbar$ found by Gao and Zhan in the context of \textsuperscript{87}Rb cold atom recoil experiments mentioned before \cite{gao-zhan}.

\subsection{Summary of bounds on iGUP parameters}

A summary of the discussed bounds on iGUP parameters is provided in Table \ref{table:bounds}. Sources marked by an asterisk provide upper limits from estimated experimental sensitivities and inputs, not actual experimental data. The table is also divided into five parts:
\begin{itemize}
\item The first lists bounds derived under particular assumptions regarding composite systems. There, the first three assume a macroscopic test body is subjected to the same modified commutator a point particle would be with same parameter $\alpha_2$ (commonly identified as $\beta$) --- an hypothesis recently challenged \cite{composite}; and the last assumes to each particle of the composite system corresponds a different parameter.
\item The first and second parts refer to constraints on models with no predefined relationship among the parameters (KMM's and Kempf's models), while the third refers to those assuming a relationship; namely, these consider $\alpha_2=2\alpha_1^2$ (ADV's model).
\item The fourth refers to bounds of gravitational origin, where sensible classical limit to GUP commutators is proposed.
\item Finally, the last part refers to bounds coming from comparison of iGUP parameters to the isotropic part of SME coefficients controlling Lorentz symmetry violation.
\end{itemize}

\begin{table*}[htb]
\centering
\caption{Upper limits on the absolute value of iGUP parameters $\alpha_n$ and $\breve{a}_n+\breve{b}_{n-2}$ in units of $(\ell_P/\hbar)^n$ derived or estimated from a variety of systems.}
\begin{tabular}{c|c c c c l}
\hline\hline
Source & $\alpha_1$ & $\alpha_2$ & $\breve{a}_1 +\breve{b}_{-1}$ & $\breve{a}_2 +\breve{b}_{0}$ & Ref.\ \TT\BB \\ \hline\hline
Sapphire split-bar mechanical resonator & - & $1.7\times10^6$ & - & $5.2\times10^6$ & \cite{bushev} \TT \\
Micro and nano mechanical oscillators & - & $1\times10^{7}$ & - & $3\times10^{7}$ & \cite{bawaj2005} \\
Gravitational wave bar detectors & - & $1\times10^{33}$ & - & $3\times10^{33}$ & \cite{marin} \\
Hydrogen 1S-2S energy difference & - & $2.23\times10^{34}$ & - & - & \cite{quesne-tkachuk10} \BB \\
\hline
Scanning tunneling microscope* & - & $\sim10^{21}$ & - & $\sim10^{21}$ & \cite{das-vagenas08} \TT \\
Lamb shift on hydrogen $s$ levels & - & $\sim10^{37}$ & - & $\sim10^{37}$ & \cite{stetsko} \\
Cold atom recoil experiment & - & $1.3\times10^{39}$ & - & $3.9\times10^{39}$ & \cite{gao-zhan} \\
Landau levels for the electron* & - & $\sim10^{50}$ & - & - & \cite{das-vagenas08} \BB \\
\hline
Accuracy on Lamb shift measurements*  & $\sim10^{10}$ & $\sim10^{20}$ & - & - & \cite{ali-das-vagenas11} \TT \\
Cold atom recoil experiment* & $2.4\times10^{14}$ & $1.2\times10^{31}$ & $4.8\times10^{14}$ & $2.3\times10^{29}$ & \cite{gao-zhan} \\
Ground state of the charmonium* & $\sim10^{17}$ & $\sim10^{34}$ & $\sim10^{17}$ & $\sim10^{34}$ & \cite{ali-das-vagenas11} \\
Landau levels for the electron* & $\sim10^{21}$ & $\sim10^{42}$ & - & - & \cite{ali-das-vagenas11} \BB \\
\hline
Binary pulsar PRS B 1913+16 data & - & $2\times10^{70}$ & - & $5\times10^{70}$ & \cite{scardigli-casadio} \TT \\
Mercury's perihelion precession & - & $2.5\times10^{71}$ & - & $7.5\times10^{71}$ & \cite{scardigli-casadio} \\
Light deflection by the Sun & - & $4.3\times10^{77}$ & - & $1.3\times10^{78}$ & \cite{scardigli-casadio} \BB \\
\hline
Torsion pendulum & - & $\sim10^{51}$ & - & $\sim10^{-51}$ & \cite{gup-sme}  \TT \\
Hydrogen 1S-2S energy difference & - & $\sim10^{30}$ & - & - & \cite{nigup} \BB \\
\hline\hline
\end{tabular}
\label{table:bounds}
\end{table*}

%%%%%%%%%%%%%%%%%%%%%%%%%
\section{Conclusions}
\label{sec:conclusions}
%%%%%%%%%%%%%%%%%%%%%%%%%

The existence of a fundamental length scale in nature is a prediction shared by many models soughing for a quantum theory of gravity. Despite theoretical and experimental efforts in the past decades  to investigate consequences of this prediction, there has been no common framework for a systematic investigation. In this work we provided such framework in the context of nonrelativistic quantum mechanics, which we dubbed the isotropic generalized uncertainty principle (iGUP) framework, constructed under the hypothesis that there is a fundamental length scale and a few other reasonable technical requirements. It was explicitly shown the iGUP framework encompasses popular proposals predicting isotropic modifications to the Heisenberg uncertainty principle related to a fundamental length scale, and by construction it is expected to encompass any such proposal indeed. Commonly cited bounds on three often investigated proposals were translated into bounds on iGUP parameters, providing an example of how the iGUP framework provides a common ground for investigation on this subject.

On the theoretical side, the iGUP framework provides a comprehensive playground for GUP model building and general investigations. Expressions for the position and momentum operators were derived as well as for commutators derived from them. In particular, the isotropic commutator $\commij=\ih[f\dij +g\hpi\hpj]$ is well-known to the literature, but to the best of our knowledge this is the first time there is a guidance to what is the allowed structure of $g(\hp)$; namely, we found $g=b_{-1}\hp^{-1} + b_0 + b_1\hp +b_2\hp^2+\cdots$. We also explored the connection of the simple one-dimensional GUP prototype $[\hx,\hp]=\ih\mathcal{F}$ to its three-dimensional extensions $\commij=\ih[f\dij+(\mathcal{F}-f)\hpi\hpj/\hp^2]$, for which we derived an interesting result: while there are infinitely many three-dimensional extensions that are noncommutative on position operators, there is at most one commutative extension and it corresponds to the case $f=\hp/(\int\mathcal{F}^{-1}d\hp)$ with $f(0)\equiv1$. Finally, these theoretical conclusions came along with the study of equivalent representations for $\hxi$ and $\hpi$ for commutative models, and were all essential when translating bounds on GUP models into iGUP parameters.

Regarding future investigations, we believe our discussion on experimental bounds suggests that there may be a current \textit{lack} of (1) experimental investigation on three-dimensional models, (2) experimental investigation on models with $\alpha_1\neq0$ and (3) also theoretical investigation on the general case $\alpha_2\neq2\alpha^2_1\neq0$. The iGUP framework naturally provides a suitable venue for advancing the third point, but regarding the first two it offers a common set of parameters to be experimentally constrained instead of diverse seemly disconnected GUP models. Especially suited for experimentalists, commutators expanded on powers of the momentum were provided for both noncommutative and commutative iGUP models --- \cf (\ref{P=p:commutator-xp-power-series}) and (\ref{P=p:commutator-xx-power-series}), and (\ref{X=x:commutator-xp}), respectively.

The quest for a fundamental length scale is to be set by experimentation, as goes with any other question on the knowledge frontier. We hope a systematic investigation as envisaged along the lines of the iGUP framework will provide further assistance on this and reduce the gap between theory and experiment.

%%%%%%%%%%%%%%%%%%%%%%%%%%%%%%%%%%%%%%%%%%%
%%%%%%%%%%%%%%%%%%%%%%%%%%%%%%%%%%%%%%%%%%%%
%

\bibliography{references}

\end{document}